\documentclass[acmsmall,screen]{acmart}

\usepackage[acronym]{glossaries}
\usepackage{graphicx}
\usepackage{textcomp}
\usepackage{xcolor}
\usepackage{float}
\usepackage{subcaption}
\usepackage{natbib}
\usepackage{xspace}   
\usepackage{booktabs}
\usepackage{hyperref}
\usepackage{balance}
\usepackage{listings}
\usepackage{multirow}
\usepackage{enumitem}
\usepackage{algorithm}
\usepackage{algorithmic}
\usepackage{makecell}
\usepackage{array}
\usepackage{tabularx}

\usepackage{amsmath}

\usepackage{tikz}
\usetikzlibrary{shapes.geometric, arrows.meta, positioning, fit, backgrounds,
                calc, shadows, decorations.pathreplacing}

\lstdefinestyle{puml}{
    basicstyle=\ttfamily\scriptsize,
    numbers=left, numberstyle=\tiny\color{gray}, numbersep=5pt,
    xleftmargin=14pt, framexleftmargin=14pt,
    keywordstyle=\color{cohA}\bfseries,
    commentstyle=\itshape\color{gray},
    morecomment=[l]{'},
    breaklines=true, showstringspaces=false, captionpos=b,
    aboveskip=4pt, belowskip=2pt
}

\newcolumntype{R}{>{\raggedleft\arraybackslash}X}
\newcolumntype{L}{>{\raggedright\arraybackslash}X}

\newcommand{\luca}[1]{\textcolor{blue}{\textbf{Luca:}~#1}\xspace}

\newcommand{\Workflow}{A-SAW\xspace}

\newcommand{\catalog}{\texttt{Item\_List.md}\xspace}
\newcommand{\skillcmd}{\texttt{/new-architecture-doc}\xspace}

\definecolor{cohA}{HTML}{4C72B0}
\definecolor{cohB}{HTML}{DD8452}
\definecolor{cohC}{HTML}{55A868}
\definecolor{regred}{HTML}{C44E52}

\newcommand{\rqonetext}{How does agentic, repository-grounded generation change the measured quality of architecture design documentation relative to manual authoring?\xspace}
\newcommand{\rqtwotext}{Which documentation-quality properties does the agent close, and which remain human-dependent?\xspace}
\newcommand{\rqthreetext}{How do practitioners experience the workflow, and how far do they trust and govern its output?\xspace}
\newcommand{\rqfourtext}{Where do the measured quality of the artifacts and the perceptions of practitioners converge, and where do they diverge?\xspace}

\newcommand{\nDocs}{45\xspace}
\newcommand{\nCohortA}{19\xspace}
\newcommand{\nCohortB}{13\xspace}
\newcommand{\nCohortC}{13\xspace}
\newcommand{\nPairs}{13\xspace}
\newcommand{\nInterviews}{13\xspace}
\newcommand{\nExploratory}{10\xspace}
\newcommand{\nRequirements}{nine\xspace}

\newcommand{\intrA}{58\%\xspace}

\newcommand{\intrC}{89\%\xspace}
\newcommand{\gainAC}{31\xspace}      

\newcommand{\sdA}{19.2\xspace}
\newcommand{\sdB}{21.0\xspace}
\newcommand{\sdC}{3.8\xspace}
\newcommand{\ciC}{[86.7, 90.7]\xspace}
\newcommand{\ciA}{[49.2, 66.1]\xspace}
\newcommand{\ciB}{[51.5, 73.3]\xspace}

\newcommand{\wilcoxW}{1.0\xspace}
\newcommand{\wilcoxP}{0.001\xspace}
\newcommand{\rankbis}{+0.97\xspace}
\newcommand{\rankbisCI}{[+0.82, +1.00]\xspace}
\newcommand{\medianGain}{+19\xspace}

\newcommand{\dOneC}{100\%\xspace}
\newcommand{\dTwoC}{93\%\xspace}
\newcommand{\dThreeC}{79\%\xspace}
\newcommand{\dFourC}{94\%\xspace}

\newcommand{\reviewImperative}{11 of 13\xspace}
\newcommand{\contextBetter}{12 of 13\xspace}   
\newcommand{\errorsFound}{9 of 13\xspace}

\newcommand{\sdRatio}{5.0\xspace}
\newcommand{\sdRatioCI}{[2.9, 9.2]\xspace}
\newcommand{\sdRatioD}{3.1\xspace}
\newcommand{\sdRatioDCI}{[1.9, 6.1]\xspace}
\newcommand{\flignerP}{0.012\xspace}
\newcommand{\flignerPD}{0.028\xspace}
\newcommand{\sdAthree}{22.9\xspace}
\newcommand{\sdCthree}{7.4\xspace}

\newcommand{\nPairsA}{nine\xspace}
\newcommand{\nPairsB}{four\xspace}
\newcommand{\srcMean}{70.5\xspace}
\newcommand{\pairsAp}{0.008\xspace}
\newcommand{\pairsAmedian}{+22\xspace}

\usepackage[absolute,overlay]{textpos}
\usepackage{tikz}
\usepackage{xcolor}
\usepackage{calc}

\newcounter{findingctr}

\newcommand{\finding}[2]{\refstepcounter{findingctr}\emph{#1:\label{box:#2}}}

\newlength{\boxw}
\newlength{\boxh}
\newlength{\shadowsize}
\newlength{\boxroundness}
\newlength{\tmpa}
\newsavebox{\shadowblockbox}

\newenvironment{findingenv}[2]%
{\vspace{0.2cm}\noindent
\begin{lrbox}{
\shadowblockbox
}
\begin{minipage}{.98\columnwidth}
\finding{#1}{#2}~}%
{\end{minipage}\end{lrbox}%
\settowidth{\boxw}{\usebox{\shadowblockbox}}%
\settodepth{\tmpa}{\usebox{\shadowblockbox}}%
\settoheight{\boxh}{\usebox{\shadowblockbox}}%
\addtolength{\boxh}{\tmpa}%
\begin{tikzpicture}
\addtolength{\boxw}{\boxroundness * 2}
\addtolength{\boxh}{\boxroundness * 2}

\foreach \x in {0,.05,...,1}
{
\setlength{\tmpa}{\shadowsize * \real{\x}}
\fill[xshift=\shadowsize - 1pt,yshift=-\shadowsize + 
1pt,black,opacity=.04,rounded corners=\boxroundness] 
(\tmpa, \tmpa) rectangle +(\boxw - \tmpa - \tmpa, \boxh - \tmpa - 
\tmpa);
}

\filldraw[fill=white!50, draw=black!80, rounded corners=\boxroundness] (0, 
0) rectangle (\boxw, \boxh);
\draw node[xshift=\boxroundness,yshift=\boxroundness,inner sep=0pt,outer 
sep=0pt,anchor=south west] (0,0) {\usebox{\shadowblockbox}};
\end{tikzpicture}\vspace{0cm}%
}

\setcopyright{none}
\begin{document}

\title{\Workflow: The Agentic Software Architecture Workflow}

\author{Finn Klatt}
\email{finn.klatt@tutanota.com}
\orcid{0009-0007-9621-1492}
\affiliation{
	\institution{University of Liechtenstein}
	\streetaddress{Fürst-Franz-Josef-Strasse}
	\city{Vaduz}
	\postcode{9490}
	\country{Liechtenstein}
}

\author{Luca Di Grazia}
\email{work@lucadigrazia.com}
\orcid{0000-0002-5306-8645}
\affiliation{
	\institution{University of St.\texorpdfstring{\,}{ }Gallen}
	\streetaddress{Torstrasse 25}
	\city{St.\texorpdfstring{\,}{ }Gallen}
	\postcode{9000}
	\country{Switzerland}
}

\author{Johannes Schneider}
\email{johannes.schneider@uni.li}
\orcid{0000-0002-3225-5564}
\affiliation{
	\institution{University of Liechtenstein}
	\streetaddress{Fürst-Franz-Josef-Strasse}
	\city{Vaduz}
	\postcode{9490}
	\country{Liechtenstein}
}

\begin{abstract}
AI assistance in software engineering focuses on implementation and verification, leaving the design phase, and in particular the maintenance of architecture documentation, comparatively underserved. \emph{Architecture as Code} makes architectural artifacts versionable, diffable, and machine-readable, but it does not reduce the effort of keeping them updated, so design documentation still ends up as a fragmented, author-dependent, and only partially trusted record of why a system looks the way it does.

In this paper we introduce the Agentic Software Architecture Workflow (\Workflow), a process-level artifact that extends Architecture as Code with a persistent agent layer: an IDE-integrated agent drafts architecture documents from a version-controlled C4 model, a typed element catalog, and instrumented templates. Every draft passes a mandatory human review gate before it is accepted.
We evaluate it in a large industrial organization along two tracks: a rubric assessment of \nDocs architecture documents including \nPairs matched source$\rightarrow$regeneration pairs, and \nInterviews task-anchored practitioner interviews. Our results show that the workflow collapses the between-author variance that made quality unpredictable ($SD$ \sdA $\rightarrow$ \sdC; ratio \sdRatio, 95\% CI \sdRatioCI). 
Intrinsic quality also rises, by a median of \medianGain points across the matched pairs ($p = \wilcoxP$, $r = \rankbis$). The cohort means of \intrA and \intrC bundle templates, catalog, agent, and a richer regeneration context. So they bound the agent's contribution from above rather than isolating it. In the agent-generated documents, completeness (100\%), consistency (93\%), and machine-readability (99\%) reach or approach the rubric maximum, while traceability reaches 79\%. After the hands-on task, the interviewed practitioners described their trust as conditional. They tied it to how much the agent had to infer beyond the data in the repository and to how many reliable generations they had seen. Their trust also covered whether the recorded design is sound. Because the rubric measures none of these, high structural scores do not by themselves show that a document can be trusted. Eleven of the 13 participants rejected agent output that no human had reviewed, and one warned that polished output may lead reviewers to skip the review. Our work supports keeping human review mandatory and concentrating it where the agent must infer beyond the repository.
\end{abstract}

\ccsdesc[500]{Software and its engineering~Software architectures}
\ccsdesc[500]{Software and its engineering~Software documentation}
\ccsdesc[300]{Computing methodologies~Intelligent agents}

\keywords{Software Architecture Documentation, Architecture as Code, LLM Agents,
C4 Model, Human-in-the-Loop, Design Science Research, Industrial Case Study}

\maketitle

\section{Introduction}\label{sec:intro}

Large language models have moved from autocompletion to agentic workflows, and in software engineering they are now widely adopted~\cite{zhang2023copilot,peng2023copilotproductivity,cui2025genai}. Agents now localize, reproduce, and fix bugs~\cite{yang2024sweagent}, coordinate in role-distributed teams~\cite{hong2024metagpt,wu2023autogen}, and use commercial issue trackers under engineer supervision~\cite{takerngsaksiri2025hula}. That effort is not well distributed across the life cycle: the study by \citet{hou2024llm4se} finds the majority of AI for Software Engineering approaches target code generation and software testing, and few target software design and software architecture. The asymmetry is understandable: code compiles or it does not. In contrast, architectural artifacts are communicative and partly subjective because business intent enters in the design phase. These intents are constraints later phases inherit \cite{perry1992foundations}.

In this domain, the main problem is not the generation of the software architecture, but its maintenance during the iterations and the entire software life-cycle. \citet{vanheesch2009vaporization} elaborate on \emph{architectural knowledge vaporization}: the rationale behind a design as well as its alternatives and trade-offs are the least recorded category of architectural knowledge. They evaporate as systems evolve and their designers move on. What remains diverges from the running system~\cite{herold2016degradation} and scatters across wikis, file shares, and repositories~\cite{hesse2013twinpeaks}. The \emph{Architecture as Code} (AaC) paradigm~\cite{bucaioni2025aac} addresses part of this: treating diagrams and specifications as plain-text source under version control makes architecture reviewable, diffable, and, decisively for what follows,
machine-readable. But AaC solves storage and versioning, not the maintenance problem.
A repository full of PlantUML does not update itself, flag its own drift, or synthesize rationale; \citet{bucaioni2025aac} indeed list minimizing drift and ensuring conformance among the paradigm's own open challenges. 

In this paper, we design the Agentic Software Architecture Workflow (\Workflow{}) extending AaC with a persistent agent layer between the artifact store and the human reviewer. An \emph{artifact layer} holds version-controlled, machine-readable C4 models and design documents governed by a typed element catalog that acts as a naming contract across every view; an \emph{agent orchestration layer} assembles context from that catalog, the system model, instrumented templates, and a feature-scoped context folder, and generates a structurally complete first draft; and a \emph{human review layer} approves every draft. Hallucination is constrained structurally rather than by exhortation, through five measures: catalog-grounded element references, context-folder scoping, explicit gap marking instead of plausible filler, an automated check for phantom identifiers and macro misuse, and a two-role approval gate. 

We evaluate \Workflow{} using a rubric assessment on \nDocs architecture documents across three cohorts (pre-tooling manual authoring, partial AaC adoption, and agentic generation). It exploits a matched-pairs design in which each of the \nPairs agentic documents regenerates a named human-authored source. The main result shows that the standard deviation of document scores falls from \sdA to \sdC (ratio \sdRatio, 95\% CI \sdRatioCI). Scores near the rubric maximum have little room to vary, so we also check traceability, the dimension furthest below that maximum (79\% in the agentic cohort). Its standard deviation decreases as well, from 22.9 to 7.4. Quality is therefore much more uniform across documents, and the 19-point gap between architecture and detailed design documents in the manual cohort does not recur in the agentic one. Intrinsic quality also increases, by a median of \medianGain points across the matched pairs ($p = \wilcoxP$, rank-biserial $r = \rankbis$); the cohort means of \intrA and \intrC bundle templates, catalog, agent and a richer regeneration context, and bound the agent's contribution from above rather than isolating it. \nInterviews practitioners each performed a 30-minute hands-on regeneration task and were then interviewed. They reported that the workflow cut first-draft effort from days to minutes and made design rationale easier to find. Their trust was conditional: none expressed unconditional confidence in the diagrams they had just watched being generated. Most insisted on human review (\reviewImperative), and their effort shifted from authoring toward verification.

\smallskip
As a result, the contributions of this paper are: 
\begin{itemize}
    \item \textbf{A process-level artifact:} the \Workflow, a three-layer design (artifact / agent orchestration / human review) extending Architecture as Code with agentic first-draft authoring under mandatory review, together with the element catalog as a machine-verifiable naming contract across all C4 views (Section~\ref{sec:methods}).
\item \textbf{An industrial evaluation with a matched-pairs design:} \nDocs documents across three cohorts, \nPairs of them matched source$\rightarrow$regeneration pairs, quantifying the quality gain and the variance collapse (Section~\ref{sec:rq1}).
\item \textbf{A capability boundary:} completeness, consistency, and machine-readability are within reach of a repository-grounded agent, while requirements traceability and decision auditability remain human-dependent (Section~\ref{sec:rq2}).
\item \textbf{The quality--trust divergence:} a triangulated account of why near-ceiling measured quality coexists with conditional practitioner trust, decomposed along established trust dimensions, plus five transferable design principles (Sections~\ref{sec:rq4} and~\ref{sec:dis}).
\end{itemize}

The impact of this work rests on a finding that measured artifact quality and practitioner trust come apart, because trust is gated by properties the rubric does not measure (Section~\ref{sec:rq4}). One participant pointed to the converse risk of over-trust: because the output looks polished, reviewers may relax the scrutiny the workflow depends on. Whether trust converges on measured quality under sustained use is a longitudinal question a single-session design cannot answer, and it remains a future direction.
\section{Background}\label{sec:bg}

\subsection{Architecture documentation and its quality}
\citet{clements2010documenting} establish the views-based framework together with the principles a sound description must satisfy: descriptive completeness, internal consistency across views, recorded rationale mapped to the requirements that motivated it, and usability for its intended readers. Our rubric operationalizes those four properties as its dimensions (Section~\ref{sec:evaldesign}).
Empirical work approaches the same target from the practitioner side: \citet{aghajani2019issues} taxonomize the documentation problems developers actually report, and \citet{aghajani2020practitioners} survey how practitioners rank their relevance, external corroboration that the properties we score are the ones that are relevant in practice, though the instrument itself remains ours. We generate Architecture \emph{Design} Documents (ADDs) and Detailed Design Documents (DDDs), not Architecture \emph{Decision} Records in the narrow one-record-per-decision sense of \citet{tyree2005decisions}: an ADD describes a feature's structure \emph{and} embeds its decisions, with their alternatives and rationale, in dedicated sections. In the vocabulary of \citet{iso42010}, the workflow produces and maintains an \emph{architecture description} composed of \emph{views}, with the element catalog playing the role of its \emph{correspondence rules}.

\subsection{Architecture as Code and the C4 model} \citet{bucaioni2025aac} define AaC not as "drawing diagrams with text tools" but as treating architectural artifacts with the rigor applied to source code, version control, peer review, a text-based representation, generalizing \emph{docs-as-code} from prose to the model itself. The consequence that matters here is machine-readability: unlike binary formats, PlantUML sources can be read \emph{and written} by an LLM, which is what makes agent-assisted maintenance possible at all. Within this landscape the C4 model~\cite{brown2022c4} (Context, Container, Component, Code) supplies a hierarchical abstraction whose upper levels serve as a durable system model, and following model-driven practice~\cite{silva2015mde} we treat its sources as the governing specification and the rendered diagram as a derived view. Measured against the criteria \citet{medvidovic2000adl} set for architecture description languages this notation is deliberately informal, trading analytic power for readability, low authoring friction, and legibility to language models: the workflow's objective is documentation upkeep, not formal verification.

\subsection{Agentic AI and the design phase} \citet{sch25gen} characterizes agentic AI as generative AI with stronger reasoning and interaction capabilities, enabling more autonomous behavior. Almost all realizations target implementation-phase tasks. Where generative AI has reached architecture, \citet{esposito2026genai4sa} report that rigorous verification of outputs is typically absent. They also identify that hallucination, fluent but source-unfaithful output~\cite{huang2024hallucination}, is among the most frequently reported challenges. We use \emph{architectural hallucination} for its manifestation here: a design that is internally plausible yet misaligned with the actual system. This paper distinguishes two senses of the word \emph{agentic}: In the orchestration literature~\cite{plaat2025agentic} agentic denotes a model-invocation pattern, tool calls, branching, self-correction, inside a bounded task. Our usage is at the \emph{process} level: the agent is a standing participant in an organizational design loop over a persistent version-controlled context store, under institutional review governance, while the individual act of generation is a single well-scaffolded pass. The levels-of-automation continuum of \citet{parasuraman2000levels} removes any remaining ambiguity: the workflow is high on information acquisition and analysis and deliberately low on decision selection and action implementation, sitting at roughly level~4 of their ten-point scale, "the computer suggests one alternative," which the human accepts, revises, or rejects. Retaining that decision point is precisely the "last chance" error-trapping mechanism they recommend wherever an incorrect automated decision is expensive. The workflow is therefore agentic without being autonomous, by design rather than by limitation.
\section{Our Approach}\label{sec:methods}

\subsection{Research Method and Setting}\label{sec:method}

Our research question is prescriptive rather than descriptive, not \emph{why} architecture documentation decays, but \emph{how} a workflow can be designed so that it does not. Thus, we follow Design Science Research~\cite{hevner2004dsr,peffers2007dsrm}: build a real artifact, deploy it, and evaluate it in use. The design iterated through three refinement rounds before summative measurement, and the evaluation is \emph{naturalistic}~\cite{venable2016feds}, i.e., based on real users in an organization. Because documentation drift is a known problem and its building blocks are individually established, the artifact is judged on demonstrable advance over current practice rather than conceptual novelty~\cite{gregor2013positioning}.

The setting is a large-scale industrial software product organization that adopted the C4 model as its architectural visualization standard roughly 18 months before this study and was, at its start, migrating from a repository-based modeling tool to text-based definitions with PlantUML. About 15 architects and senior developers performed design work during the period studied. The first author occupied a dual role throughout, as embedded practitioner building the artifact and as investigator collecting and analyzing evaluation data, characteristic of action-research-adjacent DSR in industry~\cite{runeson2009casestudy,checkland1998information}. It enables authentic data, tacit context, and the ability to iterate in a live setting. It increases the risk of designing toward one's own criteria and reading ambiguous evidence. Three measures reduce that risk: triangulation of interview claims against the independent rubric measurement, protocol standardization, and three exclusions from the interview pool. We excluded the researcher, the company-side supervisor (whose design involvement exceeded normal exposure, and who instead served as pilot participant), and the original author of the document used in the hands-on task, who could reconstruct it from memory rather than recover it from the artifacts.

\subsection{Problem Analysis and Requirements}\label{sec:problem}

Between October and December 2025, unstructured conversations with six Product Owners and four architects ($n = \nExploratory$) surfaced pain points in the existing workflow. These were exploratory, rough topics (no guide, no recording, no coding protocol) and served problem discovery, not evidence collection. Their themes are corroborated quantitatively by the pre-tooling rubric baseline (Section~\ref{sec:rq1}) and retrospectively by the summative interviews in June 2026. Five themes recurred: \textbf{(T1) Effort and maintenance burden:} initial authoring was manageable for experienced architects, but keeping documents current as implementations evolved was not. Without tooling to surface drift, the update decision was left to individual judgment and routinely deferred. \textbf{(T2) Findability and fragmentation:} content lived across several wiki spaces, repositories, file shares, and personal drives, with no convention for which tool was canonical for which artifact type. \textbf{(T3) Inconsistency:} templates were owned at team or individual level with no propagation mechanism -- so a reader could not predict a document's structural coverage before reading it.
\textbf{(T4) Trust deficit:} trust was author- and age-dependent, calibrated through informal mental models unavailable to newcomers.
\textbf{(T5) Context loss at handovers:} decisions lived in conversations, email, and memory rather than in the documents they informed, leaving downstream practitioners either to re-open a settled decision or to reverse-engineer intent from code. Table~\ref{tab:requirements} derives \nRequirements requirements from these themes and the measured baseline.

\begin{table}[t]
\caption{Requirements for the process artifact and the pain points they address. Figures in parentheses are Cohort~A means of the related rubric sub-criteria, as a percentage of the maximum score (Section~\ref{sec:rq1}).}
\label{tab:requirements}
\scriptsize
\begin{tabularx}{\linewidth}{@{}l L l@{}}
\toprule
\textbf{\#} & \textbf{Requirement} &
  \begin{tabular}[b]{@{}l@{}}\textbf{Addresses} (baseline \%)\end{tabular} \\
\midrule
R1 & A single, centrally accessible canonical location under version control & T2 \\
R2 & A centrally managed, uniformly versioned template structure & T3 \\
R3 & Template coverage enforced by tooling at generation time, not author discipline & T3 (52\%) \\
R4 & A shared naming authority governing all C4 identifiers across documents and views & T3 (55\%, 51\%) \\
R5 & Design decisions as a mandatory, structured section with named alternatives and rejection rationale & T5 (41\%) \\
R6 & Links to traceable requirement references (tickets or formal FR/NFR statements) & T5 (38\%) \\
R7 & AI-assisted first-draft generation reducing the manual effort of complete documents & T1 \\
R8 & Human expert review as a mandatory gate before any agent-generated artifact is accepted & T4 \\
R9 & Machine-readable artifacts in a format AI agents can parse & T1 (74\%) \\
\bottomrule
\end{tabularx}
\end{table}

\subsection{Conceptual Model: Adding an Agent Layer to Architecture as Code}\label{sec:model}

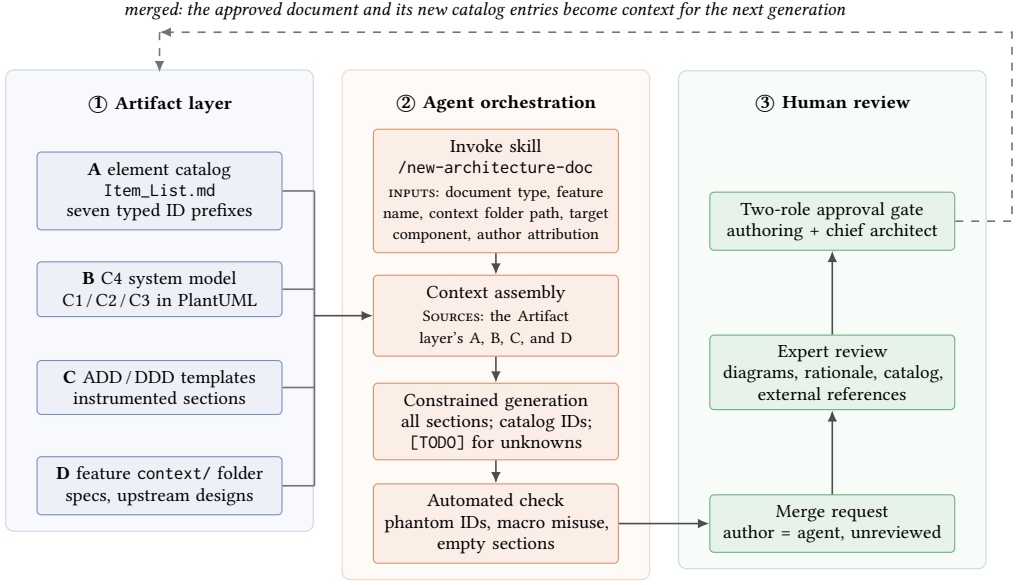
\begin{figure}[t]
\centering
\begin{tikzpicture}[
  font=\scriptsize,
  b/.style={draw,rounded corners=1.6pt,align=center,text width=31mm,inner ysep=1.3mm,inner xsep=0.7mm,execute at begin node={\hyphenpenalty=10000\exhyphenpenalty=10000}},
  art/.style={b,fill=cohA!10,draw=cohA!80}, agt/.style={b,fill=cohB!14,draw=cohB!90}, hum/.style={b,fill=cohC!14,draw=cohC!90},
  hdr/.style={font=\scriptsize\bfseries,align=center,text width=34mm},
  fl/.style={-{Latex[length=1.5mm,width=1.3mm]},semithick,black!70}, bus/.style={semithick,black!55},
  fb/.style={-{Latex[length=1.5mm,width=1.3mm]},semithick,black!55,dashed}]
\def\ca{0}\def\cb{4.45}\def\cc{8.9}

\node[hdr] (hA) at (\ca,0) {\textcircled{\scriptsize 1}~Artifact layer};
\node[hdr] (hB) at (\cb,0) {\textcircled{\scriptsize 2}~Agent orchestration};
\node[hdr] (hC) at (\cc,0) {\textcircled{\scriptsize 3}~Human review};

\node[art] (A1) at (\ca,-1.15) {\textbf{A}~element catalog \catalog\\[-0.2mm]seven typed ID prefixes};
\node[art] (A2) at (\ca,-2.45) {\textbf{B}~C4 system model\\C1\,/\,C2\,/\,C3 in PlantUML};
\node[art] (A3) at (\ca,-3.75) {\textbf{C}~ADD\,/\,DDD templates\\instrumented sections};
\node[art] (A4) at (\ca,-5.05) {\textbf{D}~feature \texttt{context/} folder\\specs, upstream designs};

\node[agt] (G1) at (\cb,-1.15) {Invoke skill \skillcmd\\[0.3mm]{\tiny\textsc{inputs:} document type, feature name, context folder path, target component, author attribution}};
\node[agt] (G2) at (\cb,-2.80) {Context assembly\\[0.3mm]{\tiny\textsc{Sources:} the Artifact layer's A, B, C, and D}};
\node[agt] (G3) at (\cb,-4.20) {Constrained generation\\all sections; catalog IDs;\\\texttt{[TODO]} for unknowns};
\node[agt] (G4) at (\cb,-5.55) {Automated check\\phantom IDs, macro misuse,\\empty sections};

\node[hum] (H1) at (\cc,-5.55) {Merge request\\author = agent, unreviewed};
\node[hum] (H2) at (\cc,-3.55) {Expert review\\diagrams, rationale, catalog,\\external references};
\node[hum] (H3) at (\cc,-1.55) {Two-role approval gate\\authoring + chief architect};

\begin{scope}[on background layer]
  \node[fill=cohA!4,draw=cohA!30,rounded corners=3pt,inner sep=2.1mm,fit=(hA)(A1)(A4)] (AL) {};
  \node[fill=cohB!4,draw=cohB!35,rounded corners=3pt,inner sep=2.1mm,fit=(hB)(G1)(G4)] {};
  \node[fill=cohC!4,draw=cohC!35,rounded corners=3pt,inner sep=2.1mm,fit=(hC)(H1)(H3)] {};
\end{scope}

\draw[bus] (2.05,-1.15)--(2.05,-5.05);
\draw[bus] (A1.east)--(2.05,-1.15); \draw[bus] (A2.east)--(2.05,-2.45); \draw[bus] (A3.east)--(2.05,-3.75); \draw[bus] (A4.east)--(2.05,-5.05);
\draw[fl] (AL.east |- G2.west)--(G2.west);

\draw[fl] (G1)--(G2); \draw[fl] (G2)--(G3); \draw[fl] (G3)--(G4);
\draw[fl] (G4)--(H1); \draw[fl] (H1)--(H2); \draw[fl] (H2)--(H3);

\draw[fb] (H3.east)--++(0.75,0)|-(11.05,0.95)--(\ca,0.95);
\draw[fb] (\ca,0.95)--(\ca,0.42);
\node[font=\scriptsize\itshape,anchor=south] at (4.3,1.0) {merged: the approved document and its new catalog entries become context for the next generation};
\end{tikzpicture}

\caption{Overview of \Workflow. The artifact layer is a version-controlled architecture repository; the agent orchestration layer reads it on demand and emits a structurally complete first draft; the human review layer is a mandatory gate no draft bypasses. Approved output re-enters the artifact layer, so the context available to the next generation grows monotonically.
}
\Description{Block diagram of three columns connected left to right. The artifact layer (left) lists four version-controlled inputs: the element catalog with typed identifier prefixes, the C4 system model in PlantUML, the ADD and DDD templates, and the feature context folder. All four feed the agent orchestration layer (middle), which runs four steps in sequence: skill invocation, context assembly, constrained generation, and an automated check for phantom identifiers, macro misuse, and empty sections. Its output enters the human review layer (right) as an unreviewed merge request, then expert review, then a two-role approval gate. A dashed arrow returns from the approval gate to the artifact layer, showing that the approved document and its new catalog entries become context for the next generation.}
\label{fig:overview}
\end{figure}

Architecture-as-Code makes architectural artifacts reviewable, diffable, and reproducible, but they remain \emph{passive}: they record the state of architectural thinking at the moment they were written and decay from that moment forward. The \Workflow{} introduces a persistent agent layer between the artifact store and the human reviewer. This turns the repository from a place where humans commit documents into the authoritative context source from which an agent reads constraints, synthesizes documentation, and surfaces inconsistencies. Figure~\ref{fig:overview} shows the three layers and the cycle connecting them.

The \emph{artifact layer} holds version-controlled documents and diagrams in a feature-scoped structure. Documents are machine-readable (Markdown with embedded PlantUML), follow standardized templates, and reference every element by a catalog identifier shared across all views; this layer is the ground truth, what humans read, review, and approve. The \emph{agent orchestration layer} is IDE-integrated and operates on the artifact layer through repository-resident skill files, complemented by separate prompt, instruction, and agent files (distinct file types serving different roles); agents are invoked on demand, never commit to the main branch, and only propose. The \emph{human review layer} gates every proposal: reviewers validate catalog accuracy, assess decision rationale, supply external references the agent cannot reach, and sign the approval table. Authorship accountability stays human; the agent supplies first-draft coverage and structural consistency. The structure attacks knowledge vaporization~\cite{vanheesch2009vaporization} by making rationale-carrying documentation the \emph{default output} of the design workflow rather than an optional retrospective effort. It produces as a by-product exactly the structured, machine-readable context whose absence drives hallucination in retrieval-based systems~\cite{zhao2024rag}. Retrieval over the repository is neither implemented nor planned here, but the artifact layer maintains its substrate.

\subsection{Architecture-as-Code Foundation}\label{sec:aac}

\textbf{View selection.} Three C4 views are used. The System Context (C1) and Container (C2) views are maintained at system level as the status quo of the involved systems and form the ground truth against which each feature is situated: C1 is the primary scoping input for Product Owners, C2 the primary concern of architects. The Component (C3) view is generated selectively, typically within a DDD, though it may also be generated standalone. The Code view is out of scope, consistent with \citet{brown2022c4}'s advice; where lower-level detail is needed, the DDD captures it at a more appropriate abstraction.

\textbf{Element catalog.} A single file, \catalog, is the naming authority for every architectural element. Elements carry one of seven typed identifier prefixes: \texttt{A\_} actor, \texttt{S\_} internal system, \texttt{SE\_} external system, \texttt{C\_} container, \texttt{CD\_} database, \texttt{CQ\_} queue, and \texttt{B\_} visual boundary. Every diagram references those identifiers in its macro calls rather than a free-form display name, yielding a machine-verifiable naming contract spanning all views and documents (R4). Listing~\ref{lst:catalog} shows the contract in use together with the two violation classes the post-generation check targets.

\textbf{Repository structure.} Each feature owns a folder holding its \texttt{ADD.md} or \texttt{DDD.md}, a \texttt{context/} sub-folder of pre-existing inputs (capability specifications, vendor API documents, requirements summaries, upstream design references), and its embedded PlantUML sources; system-level artifacts (the C1/C2/C3 diagrams, the catalog, the templates) live under a shared path visible to every feature branch. This instantiation deliberately \emph{consolidates} architecture artifacts in a dedicated central repository rather than co-locating them with the code they describe, as the original AaC formulation suggests: the organization's code spans many repositories, and centralization is what makes a single naming authority and a single system model possible. 

\begin{lstlisting}[style=puml,caption={Catalog-grounded PlantUML. Lines 1--3
are catalog entries; 5--6 are two flagged violation classes},label={lst:catalog},float=t]
System(S_1003, "Asset Management Application", "...")
Container(C_3001, "Asset Status Service", "Java/Spring", "...")
ContainerQueue(CQ_5002, "asset.status.changed", "Kafka", "...")
' --- flagged at review time
System_Ext(S_1002, "Billing", "...")     (*@\textcolor{regred}{$\leftarrow$ internal id, external macro}@*)
Container(C_3039, "Reporting Worker", "...")  (*@\textcolor{regred}{$\leftarrow$ no catalog entry}@*)
\end{lstlisting}

\subsection{Templates and Agentic Generation}\label{sec:generation}

\textbf{Two document types.} An \emph{ADD} describes a feature at the context and container levels and is written when a feature introduces or changes a container, an external dependency, or a cross-system interaction; its template has twenty sections, from scope and requirements to embedded C1/C2 (optionally C3) diagrams and design decisions. A \emph{DDD} refines an ADD into implementation details for one service or component in thirteen sections, adding, for example, component interaction diagrams and API specifications. Both share the same envelope, so the same traceability obligations apply at both levels. In the Design Decisions section, each entry must name at least one considered alternative and why it was rejected (R5); this operationalizes the reasoning-knowledge category \citet{vanheesch2009vaporization} adopt.

In the refinement rounds, output quality depended more on the \emph{specificity of the scaffold} than on prompt wording. Templates with only headings and brief placeholders yielded drafts that were structurally present but semantically thin, and expected-content descriptions with worked examples improved them ("list each functional requirement as a separate statement beginning with \emph{The system shall\dots}" instead of "describe the requirements"). Later rounds added input validation at invocation and a visual marker that keeps generated decisions findable in long documents.

\textbf{Invocation and context assembly.} A version-controlled skill file defines the draft/review/improve flow. The IDE assistant can select it autonomously when a request matches, or the practitioner invokes it explicitly, e.g., with the slash command \skillcmd{} from a prompt file, supplying the five inputs listed in Figure~\ref{fig:overview}. The agent then assembles context from four repository sources, in the order shown there: the element catalog as naming authority; the system-level C1/C2/C3 diagrams, which place the feature and separate internal from external systems by catalog prefix; the document template; and the feature's \texttt{context/} folder, the practitioner's main lever over what the agent knows. Generation ran through GitHub Copilot, backed predominantly by Claude Sonnet~4.6 and Opus~4.8. Because the quality controls are natural-language instructions that the model interprets, behavior depends on the backing model (Section~\ref{sec:threats}). In the evaluated configuration the agent had no access to the issue tracker or other external systems (possible via the Model Context Protocol or an API token, but not enabled), so requirement identifiers had to be in the context folder or be added by the reviewer; Section~\ref{sec:rq2} reports the consequence.

\textbf{Generation and validation rules.} The skill enforces five structural rules. Every template section must appear, with a visible \texttt{[TODO: \dots]} marker instead of silent omission or plausible filler where context is insufficient. C4 elements are referenced only by catalog identifier, and new elements receive a proposed identifier flagged for catalog addition; macro selection follows the identifier prefix. Feature-scope diagrams carry a new/modified legend, and design decisions follow the R5 structure. The same review procedure runs at authoring and again at merge-request review, because catalog correctness matters at both points. It blocks on identifiers that do not match the catalog and flags proposed identifiers without a catalog entry or internal identifiers emitted with the external-system macro; it also scans mandatory sections for unresolved placeholders. Taking over these machine-verifiable checks leaves the reviewer to judge content plausibility and the quality of the recorded rationale.

\subsection{Human-in-the-Loop Governance}\label{sec:governance}

The workflow is built on the principle that the agent produces and the human decides, the stance validated at industrial scale by \citet{takerngsaksiri2025hula}. The paper found practitioners accept agentic assistance precisely when they retain authority to review, edit, and approve every output. All agent-generated artifacts enter as merge requests. The document information table names the generating agent explicitly whenever no human expert review has occurred; in production use the field carries the practitioner's name, and signing is what transfers responsibility. The approval table specifies which roles must sign: the authoring architect, who validates diagram accuracy, decision rationale, and catalog correctness; and the chief architect or system owner, who approves consistency with the broader architecture (for a DDD, the architect owning the target container also reviews the component, data model, and API sections). The revision history records every change between draft and approval, including which reviewer requested it, producing a trail that distinguishes human refinement from the agent's first draft.

Five structural measures constrain architectural hallucination, responding directly to \citet{esposito2026genai4sa}'s finding that rigorous verification of generated architecture is typically absent: (i)~\emph{catalog-grounded generation} restricts element references to registered identifiers, removing the freedom to invent names for existing elements; (ii)~\emph{context-folder scoping} limits reading to files the practitioner placed there, so the agent can neither roam the repository nor reach external systems; (iii)~\emph{explicit gap marking} surfaces what the agent could not establish instead of hiding it behind plausible prose; (iv)~the \emph{automated post-generation check} catches the most common mechanical hallucination patterns before human review begins; and (v)~the \emph{two-role approval gate} is the last barrier no draft bypasses. Finally, non-determinism is treated as a documented property rather than a defect: the same inputs may yield different drafts, which is precisely why output is governed as a first draft requiring validation, with the proportion of human-confirmed content growing over the document's revision history.

\section{Evaluation}\label{sec:results}

We evaluate the \Workflow{} against four research questions (RQs):

\begin{itemize}[leftmargin=2.6em,itemsep=1pt,topsep=2pt]
  \item[\textbf{RQ1}] \rqonetext
  \item[\textbf{RQ2}] \rqtwotext
  \item[\textbf{RQ3}] \rqthreetext
  \item[\textbf{RQ4}] \rqfourtext
\end{itemize}

\subsection{Experimental setup}\label{sec:evaldesign}

\textbf{Corpus and cohorts.} The corpus is a census of the \nDocs eligible architecture documents from the partner organization, in three cohorts: \emph{A}, pre-tooling manual authoring (12 ADD, 7 DDD; $n = \nCohortA$); \emph{B}, partial AaC adoption under uncontrolled tooling use (6 ADD, 7 DDD; $n = \nCohortB$); and \emph{C}, unreviewed first-pass regenerations by the finalized workflow (8 ADD, 5 DDD; $n = \nCohortC$). Each document's last substantial-edit date, from commit or page history, separates A from B at February 2026. Candidates from the architecture repository and wiki spaces were screened manually for primary design artifacts that were substantive enough to score without speculation, reasonably complete, and not duplicated. A \emph{draft} label alone did not disqualify a document.

\textbf{The Cohort~C regeneration protocol.} The workflow stabilized after the active feature-authoring period, so no organic Cohort~C corpus exists. We constructed one by controlled regeneration. The \nPairs targets were selected on one structural criterion, a \texttt{context/} folder that still held the document's original inputs rather than on expected output quality to avoid cherry-picking. Each ran on a throw-away branch prepared by a three-step decontamination protocol: delete the targets, delete downstream and derived artifacts, and check the remaining \texttt{context/} files by content, which removed three more files carrying derivation markers. The agent saw only the feature's \texttt{context/} folder, the catalog, the system diagrams, and the templates; its output received no human correction before scoring and carried an attribution string marking it agent-generated and unreviewed.
Cohort~C therefore measures the ability to \emph{structure and regenerate} from already-rich context; it does not test novel design from a blank slate. The \nPairs sources (\nPairsA from Cohort~A, \nPairsB from Cohort~B) average \srcMean{} against a Cohort~A mean of \intrA, so the targets were better than the baseline average, as the criterion intended. Because the agent re-grounds each draft in the current catalog, a regeneration may narrow the scope it inherits, and it is scored against its own scope. 

\textbf{The rubric.} The rubric is grounded in \citet{clements2010documenting}, supported by \cite{bucaioni2025aac,herold2016degradation,vanheesch2009vaporization}, and has four dimensions with twelve sub-criteria (Table~\ref{tab:subcriteria}): \textbf{D1 completeness}; \textbf{D2 consistency}, internal and against the shared system model; \textbf{D3 traceability} of decisions to requirements and of their rationale; and \textbf{D4 maintainability} for later readers, human or agent. Only self-containedness among D4's three sub-criteria is intrinsic, so the intrinsic analyses report D4 under that name. Sub-criteria are scored 0--3 (absent, partial, substantial, full), each aggregating several component criteria recorded in the instrument. \emph{Intrinsic} sub-criteria are toolchain-independent and comparable across cohorts; \emph{process-state dependent} ones depend on infrastructure built during the project. They are reported separately and never enter intrinsic means, which keeps Cohort~A from being penalized for standards that did not yet exist. Inapplicable sub-criteria are dropped rather than scored zero, so per-document means are weighted; those in Table~\ref{tab:cohortmeans} therefore need not equal the unweighted sub-criterion columns of Table~\ref{tab:subcriteria}.\footnote{For RQ4 the rubric measures the \emph{structural quality} of
documentation, not the \emph{correctness} of the decisions it records.}
An LLM (Claude Sonnet~4.6) scored all \nDocs documents on 2026-06-18. LLM judges agree with human raters at rates comparable to human--human agreement~\cite{zheng2023judge} but carry documented biases~\cite{wataoka2025selfpreference,gu2024judgesurvey}; four properties of this setup bound every number below (Section~\ref{sec:threats}).

\textbf{Interviews.} Thirteen practitioners (7 architects, 6 engineers; professional experience from under one year to over three years) each took part in a $\sim$60-minute session in June 2026. It opened with a $\sim$30-minute hands-on task on a branch from which the original ADD of one shared feature had been removed: each participant generated a first-draft ADD with the workflow, then reviewed and improved it. The fixed feature maximizes comparability, and its original author was excluded. Four interview blocks followed (context availability, trust, process friction, governance), with identical anchor questions and adaptive probes~\cite{myers2007interview,hove2005interviews,runeson2009casestudy}; a full-protocol pilot resolved two ambiguities. 
Analysis used AI-assisted open coding without a pre-specified codebook (104 code instances); a researcher validated every code and quote against the recordings, and the validated codes were consolidated into cross-case display matrices and analyzed for convergence, divergence, and deviant cases~\cite{miles2014qda}. New theme clusters stopped emerging by roughly the ninth session, so saturation was approached~\cite{guest2006saturation}. Prior use and familiarity with the task feature serve as interpretive lenses below: seven participants had no prior hands-on use, three had used the workflow once or twice, and three regularly; eight had not worked with the task feature. Given the roughly 15 architects and senior developers who did design work in the period (Section~\ref{sec:method}) and the three principled exclusions, the pool approaches a census of the eligible population, which bounds selection effects but supports no statistical inference.

\textbf{Statistical analysis.} 
All tests are rank-based~\cite{siegel1988nonparametric}. The single pre-specified primary test is a paired Wilcoxon signed-rank test~\cite{wilcoxon1945} over the \nPairs source$\rightarrow$regeneration pairs, with the matched-pairs rank-biserial correlation as effect size~\cite{kerby2014rankbiserial} and percentile bootstrap 95\% confidence intervals (10{,}000 resamples, fixed seed) in SciPy~\cite{virtanen2020scipy}. With one rating per document, these intervals omit rater variance and are narrower than a fully specified interval would be. Dispersion is the ratio of cohort standard deviations under the same bootstrap, cross-checked with Fligner--Killeen and repeated on the dimension furthest below the rubric maximum, so that no variance claim rests only on scores bunched at the maximum. Per-dimension tests are exploratory and would need multiplicity correction~\cite{holm1979,benjamini1995fdr} to count as confirmatory.

\subsection{RQ1: How Much Does Documentation Quality Change?}\label{sec:rq1}

\begin{table}[t]
\caption{Intrinsic quality by cohort: per-dimension means and the distribution of the per-document intrinsic mean (percentage of maximum). Confidence intervals are percentile bootstrap intervals over 10{,}000 resamples.}
\label{tab:cohortmeans}
\scriptsize
\setlength{\tabcolsep}{4.5pt}
\begin{tabular}{@{}lrrrrrrcrrrr@{}}
\toprule
& & \multicolumn{4}{c}{\textbf{Dimension means}} & \multicolumn{6}{c}{\textbf{Intrinsic mean per document}} \\
\cmidrule(lr){3-6}\cmidrule(l){7-12}
\textbf{Cohort} & \textbf{n} & D1 & D2 & D3 & D4 & \textbf{mean} & \textbf{95\% CI} & \textbf{SD} & \textbf{med} & \textbf{IQR} & \textbf{min} \\
\midrule
A: baseline & 19 & 78 & 55 & 44 & 77 & \textbf{57.9} & \ciA & 19.2 & 59 & 15.5 & 16 \\
B: partial  & 13 & 79 & 51 & 53 & 79 & \textbf{62.7} & \ciB & 21.0 & 71 & 26.0 & 25 \\
C: agentic  & 13 & 100 & 93 & 79 & 94 & \textbf{88.8} & \ciC & \textbf{3.8} & 89 & 4.0 & 81 \\
\bottomrule
\end{tabular}
\end{table}

\begin{figure}[t]
\centering
\begin{subfigure}[b]{0.48\linewidth}
  \centering
  \includegraphics[scale=0.8]{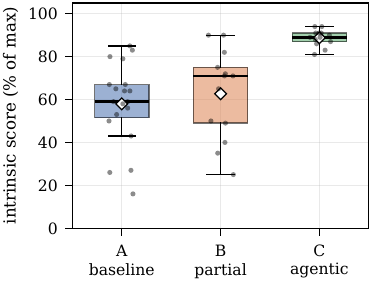}
  \caption{Distribution by cohort.}
  \label{fig:box}
\end{subfigure}\hfill
\begin{subfigure}[b]{0.48\linewidth}
  \centering
  \includegraphics[scale=0.8]{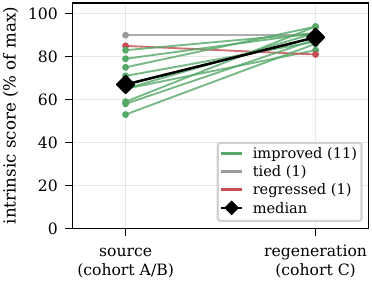}
  \caption{The \nPairs matched pairs.}
  \label{fig:slope}
\end{subfigure}
\caption{Intrinsic documentation quality ($n = \nCohortA$, \nCohortB, \nCohortC). Boxes show median and interquartile range, whiskers the $1.5\times$IQR range, white diamonds the mean, black dots individual documents. In (b) each line is one document regenerated from a named Cohort~A/B source; black diamonds mark the group medians ($67 \to 89$).}
\Description{Two panels. Panel (a) is a box plot of the per-document intrinsic score, as a percentage of the rubric maximum, for the three cohorts. Cohorts A and B sit near 58 and 63 percent with wide spread and low outliers below 25 percent; Cohort C sits near 89 percent in a very narrow box whose range does not overlap either baseline. Panel (b) is a slope chart with two vertical axes, source on the left and regeneration on the right, and one line per matched pair. Eleven lines rise, one is flat, and one falls slightly; the group median moves from 67 to 89 percent.}
\label{fig:quality}
\end{figure}

Table~\ref{tab:cohortmeans} and Figure~\ref{fig:box} report the cohort-level picture. The intrinsic mean rises from \intrA to \intrC, a gain of \gainAC points. The intermediate cohort gains only five, and unevenly: completeness(D1), maintainability(D4) roughly remain constant, traceability(D3) improves while consistency(D2) is lower. This points to partial adoption in which authors work with C4 conventions but without enforcement. Cohort~B is not a template-only condition, however: its tooling use was variable and unrecorded, and at 10\% on catalog compliance (Table~\ref{tab:subcriteria}) it had no working naming contract. Its five points therefore isolate no single component, and the \gainAC-point A$\rightarrow$C figure remains the combined effect of templates, catalog, agent, and richer regeneration context.

The more robust result is dispersion. The intrinsic-mean standard deviation falls from \sdA and \sdB to \sdC, a Cohort~A to Cohort~C ratio of \sdRatio (bootstrap 95\% CI \sdRatioCI; Fligner--Killeen $p = \flignerP$), and the interquartile range from 15.5 and 26.0 to 4.0, with Cohort~C's interval \ciC narrow and disjoint from both baselines. The collapse is not an artifact of the bounded scale: on traceability, still some 21 points below the ceiling in Cohort~C, the standard deviation falls from \sdAthree to \sdCthree (ratio \sdRatioD, 95\% CI \sdRatioDCI; $p = \flignerPD$). Quality becomes \emph{consistent rather than author-dependent}, and the same collapse closes a structural gap between document types: DDDs trailed ADDs by 19 points in Cohort~A (46 vs.\ 65) and by 13 in Cohort~B (57 vs.\ 70), whereas in Cohort~C the two are indistinguishable (90 vs.\ 88, $SD$~3.9 each).

\begin{table}[t]
\caption{Paired Wilcoxon signed-rank results over the \nPairs source$\rightarrow$regeneration pairs. Effect size is the matched-pairs rank-biserial correlation with a bootstrap 95\% CI. $^{\dagger}$pre-specified primary test; the per-dimension rows are exploratory, and all except D4 survive Holm correction at $\alpha = 0.05$.}
\label{tab:wilcoxon}
\scriptsize
\setlength{\tabcolsep}{5pt}
\begin{tabular}{@{}lccrrcc@{}}
\toprule
\textbf{Metric} & \textbf{+\,/\,--\,/\,tie} & \textbf{median src\,$\to$\,C} & \textbf{$W$} & \textbf{$p$} & \textbf{$r$} & \textbf{95\% CI} \\
\midrule
Intrinsic mean$^{\dagger}$ & 11\,/\,1\,/\,1 & $67 \to 89$ & 1.0 & \textbf{0.0010} & \textbf{+0.974} & [+0.82, +1.00] \\
D1 Completeness            & 9\,/\,0\,/\,4  & $94 \to 100$ & 0.0 & 0.0039 & +1.000 & [+1.00, +1.00] \\
D2 Consistency             & 12\,/\,1\,/\,0 & $62 \to 95$ & 5.0 & 0.0024 & +0.890 & [+0.60, +1.00] \\
D3 Traceability            & 10\,/\,3\,/\,0 & $53 \to 80$ & 6.0 & 0.0034 & +0.868 & [+0.54, +1.00] \\
D4 Self-containedness      & 5\,/\,3\,/\,5  & $89 \to 100$ & 12.0 & 0.461 & +0.333 & [$-$0.43, +1.00] \\
\bottomrule
\end{tabular}
\end{table}

The matched-pairs structure is the strongest design the corpus affords: each Cohort~C document regenerates a named source, so a within-pair test holds document identity, feature complexity, and document type fixed as no between-cohort comparison can (Table~\ref{tab:wilcoxon}, Figure~\ref{fig:slope}), though it holds constant neither the context available at authoring time nor architectural scope (Section~\ref{sec:evaldesign}). The intrinsic mean improves in 11 of 13 pairs, ties in one, and regresses in one by four points on an already-strong source: $W = \wilcoxW$, $p = \wilcoxP$, a near-maximal $r = \rankbis$ (95\% CI \rankbisCI), median improvement \medianGain points (95\% CI $[+12, +26]$). It holds on the \nPairsA Cohort~A pairs alone ($p = \pairsAp$, median \pairsAmedian{} points), so it does not rest on the \nPairsB pairs whose sources had already seen partial tooling. Three dimensions are individually significant at $p < 0.005$; only self-containedness is not, as expected from a source median of 89 that leaves little headroom and produces five ties. Gains concentrate where sources were weakest, the three lowest-scoring DDD originals gaining $+35$, $+30$, and $+33$ points, while the tie and the regression occur on strong sources: a ceiling-bounded process that lifts weak documents to the plateau strong ones already occupied.

\begin{findingenv}{Answer to RQ1}{rq1}
Intrinsic documentation quality averages 58\% in Cohort A and 89\% in Cohort C. Across the 13 matched pairs, median scores rise from 67\% to 89\%, with a median within-pair improvement of 19\% (p = 0.001, r = +0.97).
\end{findingenv}

\subsection{RQ2: Where Does the Agent's Reach End?}\label{sec:rq2}

\begin{figure}[t]
\centering
\includegraphics[scale=0.8]{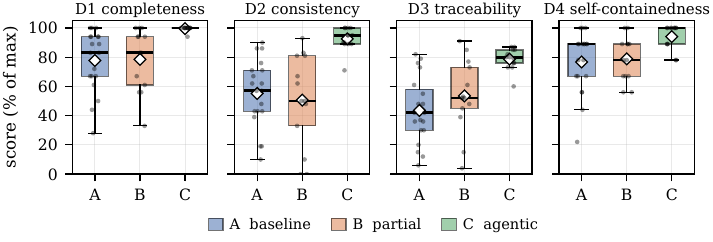}
\caption{Per-dimension intrinsic distributions by three cohorts; encoding as in Figure~\ref{fig:quality}, boxes show median and interquartile range, whiskers the $1.5\times$IQR range, white diamonds the mean, black dots individual documents. 
}
\Description{Four side-by-side box-plot panels, one per rubric dimension, each showing cohorts A, B, and C. In every panel Cohort C sits higher and far narrower than both baselines. Completeness rises to the rubric maximum; consistency has the widest baseline spread and the sharpest tightening under the workflow; traceability remains the lowest scoring dimension in all three cohorts, reaching about 79 percent in Cohort C; self-containedness starts high and gains least.}
\label{fig:dims}
\end{figure}


\begin{table}[t]
\caption{Per-sub-criterion trajectory across cohorts (\% of maximum).}
\label{tab:subcriteria}
\scriptsize
\setlength{\tabcolsep}{5pt}
\begin{tabular}{@{}llrrrrr@{}}
\toprule
\textbf{Dependence} &\textbf{Sub-criterion} &\textbf{A} &\textbf{B} &\textbf{C} &\textbf{A$\to$B} &\textbf{B$\to$C} \\
\midrule
\multirow{8}{*}{Intrinsic}
& 1.2 C4 element completeness    & 73 & 67 & 100 & $-$6  & $+$33 \\
& 1.3 Content depth              & 83 & 91 & 99  & $+$8  & $+$8  \\
& 2.1 C4 alignment               & 61 & 53 & 92  & $-$8  & $+$39 \\
& 2.3 Structural consistency     & 51 & 49 & 94  & $-$2  & $+$45 \\
& 3.1 Requirements linkage       & 38 & 50 & 87  & $+$12 & $+$37 \\
& 3.2 Implementation linkage     & 55 & 63 & 78  & $+$8  & $+$15 \\
& 3.3 Decision auditability      & 41 & 51 & 72  & $+$10 & $+$21 \\
& 4.3 Self-containedness         & 77 & 79 & 94  & $+$2  & $+$15 \\
\midrule
\multirow{4}{*}{Process}
& 1.1 Template adherence         & 52 & 57 & 92  & $+$5  & $+$35 \\
& 2.2 Catalog compliance         & 55 & 10 & 72  & $-$45 & $+$62 \\
& 4.1 Format consistency         & 86 & 94 & 100 & $+$8  & $+$6  \\
& 4.2 Machine-readability        & 74 & 58 & 99  & $-$16 & $+$41 \\
\bottomrule
\end{tabular}
\end{table}

Table~\ref{tab:subcriteria} and Figure~\ref{fig:dims} localize the gain, and its non-uniformity is the answer to RQ2.

\textbf{What the agent closes.} Completeness reaches its ceiling, C4 element completeness at 100\% and content depth at 99\%. Consistency, the dimension with the widest baseline spread and the sharpest agentic tightening, shows the largest intrisic gains from Cohort B to Cohort C: structural consistency $+45$ points from Cohort~B and C4 alignment $+39$. The reason is that macro rules and boundary conventions are enforced mechanically rather than by author discipline. Machine-readability raises from a mid-adoption of 58\% to 99\% since generation always emits valid includes, correct macro syntax, and standard Markdown. Consistency is also the gain most clearly attributable to the \emph{agent}: validating each element against a shared catalog is work a template cannot perform on its own. The completeness and format gains could largely be reproduced by a diligent human filling the same instrumented template, so those conflate template enforcement with agent capability.

\textbf{What it does not close.} Traceability remains the lowest scoring dimension even under the workflow, at \dThreeC and the only one short of the ceiling, its weakest sub-criteria being decision auditability (72\%) and implementation linkage (78\%). Table~\ref{tab:gaps} classifies the residual gaps by root cause, and the classification matters more than the numbers: none is a first-draft \emph{drafting} failure. Some are configuration artifacts, some structural properties of one-pass generation, and three are workflow gaps a write-back step and two assertions would eliminate.

\begin{table}[t]
\caption{Root-cause classification of the residual Cohort~C quality gaps. SC = Sub-criterion. 
}
\label{tab:gaps}
\scriptsize
\begin{tabularx}{\linewidth}{@{}l l l@{}}
\toprule
\textbf{Residual gap} & \textbf{SC} & \textbf{Root cause} \\
\midrule
Ticket references missing (7 of 13 docs score 0) & 3.1 & Configuration: no issue tracker and no optional integration  \\
No evolution evidence (all 13 score 0--1) & 3.3 & Structural: a single-pass first draft has no prior draft to diff against \\
Downstream references absent (6 of 8 ADDs) & 3.2 & Context: downstream dependency data not yet in the repository \\
Phantom identifier w/o catalog entry (4 docs) & 2.2 & Workflow gap: no write-back of newly proposed identifiers \\
External macro on internal systems (4 ADDs) & 2.2 & Workflow gap: prefix rule not cross-checked against the catalog \\
Missing hyperlinked table of contents (6 docs) & 4.3 & Workflow gap: template lacks a mandatory ToC section \\
\bottomrule
\end{tabularx}
\end{table}

The decision-auditability figure deserves unpacking, because it is depressed by an artifact of the evaluation design rather than by drafting quality. Its \emph{evolution evidence} criterion scores 0--1 for all 13 Cohort~C documents by construction whereas its three substantive criteria behave very differently, decision rationale and alternatives-considered reaching the maximum in 12 of 13 documents. Without the structurally zero criterion, decision auditability would average roughly 90\% and lift the dimension from \dThreeC to about 85\%. We retain it for consistency with the full rubric, which biases traceability \emph{against} the workflow. What survives that correction is the genuine boundary: requirements linkage depends on external systems the agent was not connected to, and decision \emph{evolution} depends on iterative human authorship a generation pass cannot manufacture.

\begin{findingenv}{Answer to RQ2}{rq2}
The agent closes structural properties and leaves relational ones open.
Completeness (\dOneC), consistency (\dTwoC), and machine-readability (99\%, process-state dependent) reach or approach the ceiling given a well-maintained repository, and consistency is the gain most clearly attributable to the agent rather than the template. Traceability remains lowest (\dThreeC).
\end{findingenv}

\subsection{RQ3: How Do Practitioners Experience and Govern the Workflow?}\label{sec:rq3}

The five scale-type anchor questions distribute as follows ($n = \nInterviews$; categories collapsed from the coded responses, ambiguous ones classified by dominant content). Qualitative codes are given below. Context sufficiency distribution: sufficient (5), partial (4), insufficient due to an own-domain gap (3), sufficient with prior knowledge (1). Context access, old$\to$new: better (10), same-to-better (2), uncertain (1), worse (0). Trust in generated diagrams: high (2), medium-high (3), medium (5), medium-low (1), medium$\to$high after iteration (1), n/a due to render failure (1). Trust, old$\to$new: more (7), same (2), less (4). Concern over unreviewed output: strong (11), moderate (1), none (1).

\textbf{Context availability improves due to one key mechanism.} Ten of thirteen rated the new approach better for locating design rationale and two same-to-better (\contextBetter{} judging it at least as good), and none judged the old approach categorically superior. The mechanism they described is that the structured repository moves context assembly from manual labor to standing infrastructure: "it would take me days to do that on my own [\dots] and probably even with AI, but without all the context, the result would not be as good as this one" (P04). Two participants named the condition attached to that gain: the labor is relocated, not eliminated, to whoever curates the feature context ("that job was saved, but that also is some work," P03), and two others identified an unresolved staleness boundary, since repository-resident context can lag decisions still being negotiated elsewhere (P07, P09).

\textbf{Trust clusters at medium, and four patterns explain why.} The modal response was "medium" (5 of 13): None reported unconditional high trust, and even the two high ratings carried caveats. (i)~\emph{Trust decays with abstraction level:} four participants described a gradient in which trust is highest where the agent draws on existing structured data and lowest where it must infer across undocumented systems, "it goes from a 90 to C1 to 60, 70 to C4" (P06), llustrating lower trust at more detailed abstraction levels. (ii)~\emph{Trust is earned, not granted:} "at least for four to five designs, it should generate those documents [\dots] accurately, then only I will be able to trust it" (P01), calibrated reliance in the sense of \citet{leesee2004trust}, accruing over observed interactions rather than tracking surface plausibility. (iii)~\emph{Gains are attributed to input structure, not model capability:} "I know that the information it takes is much more structured, [so] the agent makes less errors" (P06). Consistently, all four "less trust now" responses came from the no-prior-use group, suggesting that first exposure can temporarily \emph{lower} trust by making limitations salient while sustained use rebuilds it, suggestive only at seven no-prior-use versus six experienced participants. (iv)~\emph{Trust is split by modality:} several trust the agent more for template compliance than for diagram semantics, "in terms of following a template [\dots] I would trust AI-generated output more, because humans miss some details" (P13).

The strongest convergence in the dataset concerns review: \reviewImperative stated unequivocally that an agent-generated document used without human review is unacceptable, "I will not accept [it] for consumption at all" (P08). The two exceptions prove the rule from opposite directions: one expressed no AI-\emph{specific} concern, but from a position of universal distrust of all documentation, framing the agent as a "junior architect" whose output must, like any junior's, be reviewed; the other saw the concern diminishing over time \emph{if} accuracy proves consistent. In productive tension with that consensus, P11 identified a \emph{polished-output hazard}: precisely because the document looks professional, "sometimes people try to avoid the review", the over-trust failure mode \citet{parasuraman1997misuse} term misuse, and the inverse of the low-trust problem.

\textbf{Friction: generation compresses, verification expands.} All thirteen reported faster first drafts, repeatedly quantified as a collapse from days or weeks to minutes. The same participants located the counterweight: "the document generation is simple. [\dots] To verify that document, to review that document, it's the tough task" (P01), with P03 estimating "at least a day [\dots] to read, understand, and correct it." Critically, the \emph{nature} of review changes, not only its magnitude: P01 previously reviewed documents "from the solution point of view, whether this solution will work in production" and now checks "whether the arrow is correct." The net accounting nonetheless stayed positive for most, P13 judging that the workflow "has improved the speed [\dots] even including the time to review it." A different nine accepted the output largely as-is during the session, which must not be read as endorsement: several attributed acceptance to the thirty-minute time box and to unfamiliarity with the feature ("normally [I] would review every line," P03). Adoption friction concentrated at first use, "from the second document onwards I got into the rhythm" (P08), with two boundary cases: one participant could not start without explicit instructions despite genuine interest, and downstream consumers were not yet ready to consume Markdown rather than wiki pages.

\textbf{Governance: ownership tracks intellectual contribution.} Felt ownership ranged from unambiguous personal accountability ("we control the architecture. The AI does not control the architecture. But we use AI to document that architecture," P11), through ownership conditional on reviewing and signing (P05), to its complete absence, and the last is diagnostic: "I didn't feel any ownership [\dots] because there was no input from my side. I just clicked and asked the question" (P02). Review-and-sign is what converts a generated draft into an owned artifact. \errorsFound participants identified or suspected at least one concrete error, and the four who did not all lacked both feature knowledge and review time. Errors ranged from a mis-stated publish/subscribe relationship and diagram--text drift to an ADD generated for the \emph{wrong component entirely}, exactly the architectural-hallucination mode the governance design targets. That error detection tracks expertise-plus-time so closely has a sharp consequence: the time pressure that makes the workflow attractive can suppress the very review that makes its output trustworthy.

\begin{findingenv}{Answer to RQ3}{rq3}
Practitioners rated better by 10 of 13 and at least as good by 12 of 13. \reviewImperative reject unreviewed agent output outright, independently validating the mandatory review gate. Effort shifts from authoring to verification and changes in kind from solution validation to correctness checking, and ownership tracks intellectual contribution, produced by review-and-sign rather than by generation.
\end{findingenv}

\subsection{RQ4: Where Do Measurement and Perception Diverge?}\label{sec:rq4}

Read against each other the tracks yield three convergences and one divergence, and the divergence is the most analytically valuable result.

\textbf{Three convergences.} On \emph{context}, participants found the "why" easier to recover and the rubric shows the gain (traceability 44\%~$\to$~\dThreeC, self-containedness \dFourC), and both tracks locate the residual in the same place, context staleness and the upstream curation labor the repository relocates rather than removes. On \emph{friction}, reduced-creation-friction claims co-occur with the expected process-maturity scores (format consistency 100\%, template adherence 92\%) and the live-task corrections fall into the same error classes as the residual gaps of Table~\ref{tab:gaps}; but the rubric scores the artifact, not the effort to verify it, so the most-cited process change, that generation became easy and verification hard, is invisible to it. On the \emph{human-dependent residual}, ownership proved proportional to intellectual contribution, and decision auditability, held down by absent evolution evidence, is with catalog compliance the lowest Cohort~C sub-criterion.

\textbf{The divergence.} The rubric rates Cohort~C consistency at \dTwoC and C4 element completeness at \dOneC, yet the practitioners who generated and reviewed their own drafts describe trust as conditional rather than settled. The two measurements are not taken on the same artifacts: the rubric scores the thirteen regenerations, prepared under a controlled protocol on features whose context was already rich, whereas the trust reports concern session drafts that were never scored, so part of the difference may be an unmeasured output-quality gap rather than a perception effect. What that rival does not account for is the \emph{structure} of what practitioners said: trust is gated by three factors the rubric does not measure (abstraction level, accumulated reliability over repeated use, and, running the other way, a complacency hazard) and by one it cannot measure, since their notion of trust extends to whether the recorded design is \emph{right}, which is architectural decision quality rather than documentation quality. The instrument does not certify the property practitioners most want certified; the discussion below decomposes these gates against established trust dimensions. Across all four pairings the tracks triangulate on one boundary: where the agent is grounded in the structured repository both register strong gains, the variance collapse mirrored in P10's account that the agent will "more or less put the same level of rigor on every part,'' and where the artifact depends on external grounding and iterative human judgement, both register the residual.

\begin{findingenv}{Answer to RQ4}{rq4}
Three of the four pairings converge; the fourth diverges systematically. Near-ceiling measured consistency (\dTwoC) coexists with conditional practitioner trust, because trust is gated by abstraction level, earned through repeated use, vulnerable to polished-output complacency, and scoped to a property the instrument does not measure. 
\end{findingenv}

\section{Discussion}\label{sec:dis}

\subsection{Structural Quality and Trust}

The divergence of Section~\ref{sec:rq4} looks paradoxical only if trust is treated as a single quantity that high quality should produce. \citet{leesee2004trust} distinguish trust in an automation's \emph{performance}, \emph{process}, and \emph{purpose}, and \citet{mcknight2002trust} draw a parallel distinction among \emph{competence}, \emph{integrity}, and \emph{benevolence} beliefs. The conditions participants attached to trust (Section~\ref{sec:rq3}) map onto these dimensions, as an interpretive alignment rather than a validated coding. The abstraction-level condition concerns performance trust, a competence belief that high consistency scores cannot establish where the agent must infer across a thin system model. Trust earned through repeated reliable use is process trust, which no single draft can supply. Whether the recorded design is \emph{right} is purpose trust, an integrity belief outside the documentation-quality construct. The polished-output hazard runs the other way and has no McKnight counterpart. It is an \emph{irony of automation}~\cite{bainbridge1983ironies}: the more fluent the automation, the less vigilantly it is monitored.

Process trust also accrues asymmetrically. People who see an algorithm err lose confidence in it faster than in a human making the same error, and stay reluctant even when it outperforms the human on average (\emph{algorithm aversion}~\cite{dietvorst2015aversion}). One visible error can therefore set back trust built over many reliable generations, which strengthens the case for a review gate that intercepts errors before they reach the practitioner. For practice, higher artifact quality alone does not establish that artifacts will be trusted and relied upon. The same irony carries a skill-degradation risk that a single-session design cannot observe~\cite{bainbridge1983ironies}: teams that rely on the agent for first drafts may lose the fluency to write architecture documentation unaided, and with it the expertise that review depends on. Review should therefore remain a genuine cognitive act.

\subsection{Transferable Design Principles}

Five principles are intended to transfer; following \citet{lincoln1985naturalistic}, the receiving context judges transferability against Sections~\ref{sec:method} and~\ref{sec:evaldesign}.
\textbf{(P1)}~\emph{Ground the agent in a persistent, version-controlled context store}, such as the central AaC repository that makes grounded first drafts possible.
\textbf{(P2)}~\emph{Enforce a machine-verifiable naming contract}: a shared, typed identifier catalog turns an informal convention into a constraint the agent satisfies by lookup instead of inference.
\textbf{(P3)}~\emph{Separate the artifact, agent-orchestration, and human-review layers}, so that automation and governance coexist and accountability sits at the approval gate.
\textbf{(P4)}~\emph{Calibrate review to the agent's competence gradient}: concentrate review where the agent must infer, and guard it against the polished-output hazard.
\textbf{(P5)}~\emph{Design for earned trust and progressive autonomy}: start with every output reviewed, and grant more autonomy along the levels-of-automation continuum~\cite{parasuraman2000levels} only as reliability is demonstrated.

P2 and P4 are \emph{empirically derived} here: P2 from the measured reduction in cross-document inconsistency, the problem \citet{szczepanik2025c4agents} independently report, and P4 from the abstraction-level condition that four of the thirteen participants stated, a thin base offered as such. P1, P3, and P5 are \emph{literature-supported}~\cite{bucaioni2025aac,takerngsaksiri2025hula,parasuraman2000levels} \emph{and consistent with our experience, but untested}. In particular, every regeneration target was selected for sufficient context, so P1's premise that output quality tracks context richness is assumed. The transferable contribution lies less in any single principle than in their composition into one governable workflow.

\subsection{Practical Implications and Boundary Conditions}

\textbf{Prerequisites and the practitioner's quality lever.} Besides the preconditions listed under boundary conditions below, the workflow needs an IDE assistant with repository-resident extension support. Building the catalog and the C1/C2 ground truth took a multi-month infrastructure phase, which is required for coherent multi-author documentation at scale, with or without an agent. After that, we expect output quality to depend above all on how complete the feature's \texttt{context/} folder is at invocation. Its preparation deserves first-class status with per-document-type guidance, and part of the saved drafting effort is likely to be re-spent there. The highest-value extension is connecting the issue tracker, whose absence accounts for 7 of the 13 agentic documents scoring zero on ticket linkage, together with a convention that links tickets to features (requirements as code, or a ticket reference in the \texttt{context/} folder), because access alone does not tell the agent which ticket belongs to which feature.

\textbf{Protect review time, and frame the case as reallocation.} Error detection required both domain expertise and dedicated review time (Section~\ref{sec:rq3}), so a workflow sold on speed can undermine its own quality assurance unless part of the saved time is reinvested. Review should be budgeted and non-optional, and it should go to reviewers with the relevant domain expertise, supported by a checklist and the mandatory catalog and macro checks. The business case should promise \emph{reallocation} rather than reduction: a faster, more consistent first draft, with review effort that may approach the drafting effort saved, or exceed it for complex features, but is spent on higher-value work.

\textbf{Boundary conditions.} We expect the quality and consistency findings to depend on structural preconditions more than on this organization: a C4 system model maintained at container level, version-controlled text-based diagrams with a shared macro library that an LLM can read and write, a typed element catalog or equivalent naming authority, and a codebase spread over enough repositories that centralization pays. Where these hold, the \emph{direction} of the quality and variance findings should recur even if the magnitude differs. The findings should \emph{not} be extended to greenfield design of novel architecture, which the evaluation did not test. In small single-repository settings, co-locating architecture with code is likely more proportionate than the central repository of Section~\ref{sec:aac}. Traceability and trust remained human-dependent in this study.
\section{Threats to Validity}\label{sec:threats}

\subsection{Construct and internal validity.} The rubric's construct boundary (Section~\ref{sec:evaldesign}) leaves ``effectiveness of the design phase'' only partly operationalized, so part of the consistency--trust gap reflects construct scope rather than a measurement discrepancy. A single rater in a single pass makes measurement error inseparable from true document variance, and without a second rating neither Cohen's $\kappa$~\cite{cohen1960kappa} nor the intraclass correlation~\cite{shrout1979icc} can be computed. The rater also shares a model family with the generator, inviting self-preference bias~\cite{wataoka2025selfpreference,gu2024judgesurvey}, and was not blind to cohort, since Cohort~C carries an agent-generated attribution string; unrecorded document length leaves verbosity an unmeasured rival for part of the level difference. Only the 0--3 behavioral anchors and triangulation against the practitioner track bound these biases. Cross-cohort comparison is associational, and even the matched pairs cannot separate the agent's contribution from the instrumented templates and the richer regeneration context. Two design effects push in opposite directions: first drafts lack revision history, which depresses decision auditability \emph{against} the workflow, whereas regenerating from existing context is easier than blank-slate design. The first author's dual role as designer and evaluator leaves residual bias in the rubric and in reading ambiguous evidence, which triangulation, protocol standardization, and the three subject-pool exclusions (Section~\ref{sec:method}) reduce but do not remove.

\subsection{External validity and reliability.} A single organization and stack at one C4 maturity level support \emph{transferability}~\cite{lincoln1985naturalistic} rather than statistical generalization, so magnitudes are feasibility evidence, not predictions. In the interview task, the time box makes in-session acceptance a poor proxy for endorsement, and the single shared feature limits coverage of feature complexity. Participants' unfamiliarity with the feature understates error detection relative to a domain expert, and prior exposure to a demonstration is a social-desirability risk that the task-first design only partly offsets. One human coder validated the AI-assisted first pass, so no inter-coder reliability statistic exists. Because the quality controls are natural-language instructions, behavior depends on the backing model and would need re-validation on another; the principles are model-agnostic, but their instantiation is not. The exploratory conversations that seeded the problem analysis were unrecorded, so their themes, though corroborated by the baseline scores and later interviews, cannot be re-analyzed independently.
\section{Related Work}\label{sec:related-work}

\subsection{Generative AI for architecture and design documentation} This is where our contribution sits. \citet{esposito2026genai4sa} find rigorous verification of generated architecture typically absent and hallucination among the most reported challenges, and call for AI integration at the process rather than the task level; we answer with a process-level artifact whose verification is structural (catalog grounding, gap marking, an automated check, a two-role gate) rather than exhortative. \citet{szczepanik2025c4agents} synthesize C4 diagrams from natural-language requirements in one shot and report that shared conversational memory cannot keep terminology consistent, whereas we maintain a multi-artifact documentation base in a production repository and our element catalog realizes the persistent naming authority their agents lacked. \citet{birru2025docupdates} update documentation from a code-change trigger and generate a patch, where ours is a practitioner invocation producing a complete first-draft design document as a repeatable organizational practice.
\citet{nicacio2022devopsdiagrams} generate diagrams from DevOps descriptors and could ensure consistency only to a limited extent; we derive from a hand-maintained architectural model and enforce consistency through an explicit identifier contract. \citet{cheng2025genai4re} find generated requirements rarely feeding downstream architectural phases, the handover discontinuity our context folder and requirements-linkage criteria target, though closing it fully remains future work.

\subsection{Architecture documentation, decisions, and Architecture as Code} Our four rubric dimensions operationalize, rather than extend, the views-based canon of \citet{clements2010documenting}; an externally grounded construct would instead come from the reported-issue taxonomy and practitioner rankings of \citet{aghajani2019issues} and \citet{aghajani2020practitioners}, so our derivation from the canon and from Section~\ref{sec:problem} is treated as a construct-validity threat in Section~\ref{sec:threats}. We target the knowledge vaporization \citet{vanheesch2009vaporization} characterize, but make rationale-carrying documentation the workflow's default output rather than a discipline the architect must sustain, and adopt the first-class treatment of decisions of \citet{tyree2005decisions} in substance without its one-record-per-decision format. Our cohort comparison adds one data point to the thin empirical base on degradation reported by \citet{herold2016degradation}, for documentation quality rather than code--architecture conformance, and where \citet{hesse2013twinpeaks} treat requirements and architecture documentation jointly we address the architectural side alone. Our agent layer attacks the drift and conformance challenges \citet{bucaioni2025aac} leave open for Architecture as Code, with the deliberate deviation of centralizing rather than co-locating artifacts, and generalizes the docs-as-code precursor of \citet{thomchick2018docsascode} from prose to the model. We adopt the C4 model of \citet{brown2022c4}, Code view excluded on its own advice, as an agent's machine-readable context source rather than the comprehension aid \citet{vazquez2020c4} evaluate; the premise of \citet{silva2015mde} that a diagram is a derived view of a specification; a notation deliberately informal against the criteria of \citet{medvidovic2000adl}; and the view/viewpoint/correspondence vocabulary of \citet{iso42010}, borrowed without claiming conformance.

\subsection{Trust, automation, and the evaluation of generated artifacts} That near-ceiling measured quality does not produce high trust we explain (Section~\ref{sec:dis}) with the performance/process/purpose decomposition of \citet{leesee2004trust} and the competence/integrity/benevolence beliefs of \citet{mcknight2002trust}; the misuse--disuse distinction of \citet{parasuraman1997misuse} names the polished-output hazard one participant surfaced; the scale of \citet{parasuraman2000levels} places the workflow at level~4; the ironies of automation of \citet{bainbridge1983ironies} we carry forward as a longitudinal risk this design cannot observe; and the asymmetric confidence drop after an observed error shown by \citet{dietvorst2015aversion} is why we argue the review gate protects accrued trust rather than only catching defects. On measurement we adopt the LLM-judge paradigm \citet{zheng2023judge} establish while treating as live threats the reliability problem \citet{gu2024judgesurvey} identify as central and the self-preference bias \citet{wataoka2025selfpreference} demonstrate, defending the scores by triangulation rather than by the paradigm's reputation. Architectural hallucination is the domain-specific case of the taxonomy of \citet{huang2024hallucination}.
\section{Conclusion}\label{sec:conclusion}

We propose \Workflow, a process-level artifact that extends Architecture as Code with a persistent agent layer: an IDE-integrated agent fills instrumented templates from a version-controlled C4 model and a typed element catalog, and every draft passes a mandatory human review gate. Across \nDocs documents and \nInterviews task-anchored interviews in a large industrial organization, intrinsic quality rose by a median of \medianGain points over \nPairs matched pairs ($p = \wilcoxP$, $r = \rankbis$), and the standard deviation of document scores fell from \sdA to \sdC, while traceability remained the lowest dimension. Practitioners based their trust on properties the rubric does not measure, so high rubric scores do not by themselves show that a document can be trusted. 

\balance
\bibliographystyle{ACM-Reference-Format}
\bibliography{references/paperbib}
\end{document}